\documentclass[preprint,12pt]{elsarticle}

\usepackage{amsmath,amssymb}
\usepackage{bm}
\usepackage{hyperref}

\newcommand\pythia{\textsc{Pythia}}
\newcommand\diag{\operatorname{diag}}
\newcommand{\covDeriv}{\partial}
\newcommand{\etap}{\eta_\mathrm{p}}
\newcommand{\etapa}{\etap^\mathrm{a}}
\newcommand{\etapb}{\etap^\mathrm{b}}
\newcommand{\sNN}{\sqrt{s_\mathrm{NN}}}
\newcommand{\pT}{p_\mathrm{T}}

\journal{Physics Letters B}

\usepackage[normalem]{ulem}  
\ifpdf
  \usepackage[pdftex]{color}
\else
  \usepackage[dvipdfmx]{color} 
\fi

\begin{document}

\begin{frontmatter}



\title{Effects of hydrodynamic and initial longitudinal fluctuations\\
on rapidity decorrelation of collective flow}

\author[label1]{Azumi Sakai}
\address[label1]{
Department of Physics, Sophia University, Tokyo 102-8554, Japan
}
\ead{a-sakai-s4d@eagle.sophia.ac.jp}
\author[label2,label3]{Koichi Murase}
\address[label2]{
Yukawa Institute for Theoretical Physics, Kyoto University, Kyoto 606-8502, Japan
}
\address[label3]{
Center for High Energy Physics, Peking University, Beijing 100871, China
}

\ead{koichi.murase@yukawa.kyoto-u.ac.jp}
\author[label1]{Tetsufumi Hirano}
\ead{hirano@sophia.ac.jp}


\begin{abstract}
We investigate the interplay between hydrodynamic fluctuations and initial longitudinal fluctuations
for their effects on the rapidity decorrelation of collective flow in high-energy nuclear collisions.
We use a (3+1)-dimensional integrated dynamical model
in which we combine initial conditions with longitudinal fluctuations,
fluctuating hydrodynamics and hadronic cascades.
We analyse the factorisation ratio
in the longitudinal direction
to study the effect of these fluctuations on the rapidity decorrelation.
We find an essential difference between the effects of the hydrodynamic fluctuations and the initial longitudinal fluctuations
in the centrality dependence of the factorisation ratios.
A combination of the hydrodynamic fluctuations
and the initial longitudinal fluctuations
leads to reproduction of the centrality dependence of the second-order factorisation ratio,
$r_2(\etapa,\etapb)$, measured by the CMS Collaboration.
Our model also qualitatively describes the centrality dependence of
the third-order factorisation ratio, $r_3(\etapa,\etapb)$.
These results demonstrate the importance of the hydrodynamic fluctuations,
as well as the initial longitudinal fluctuations,
in understanding the longitudinal dynamics of high-energy nuclear collision reactions.

\end{abstract}




\begin{keyword}
quark--gluon plasma, relativistic fluctuating hydrodynamics, factorisation ratios


\end{keyword}

\end{frontmatter}



High-energy nuclear collision experiments
performed at the Relativistic Heavy Ion Collider (RHIC) at
Brookhaven National Laboratory and the Large Hadron
Collider (LHC) at CERN aim at understanding the bulk and transport
properties of the deconfined nuclear matter, the
quark--gluon plasma (QGP)~\cite{Yagi:2005yb}.
One of the major discoveries
at RHIC
is the large magnitude of the second-order azimuthal anisotropy%
~\cite{Ackermann:2000tr,
Adler:2001nb,Adams:2003am,Adcox:2002ms,Adler:2003kt,Back:2002gz},
also known as the elliptic flow%
~\cite{Ollitrault:1992bk}.
The elliptic flow turned out to be consistent with the results
from ideal hydrodynamic models%
~\cite{Kolb:2000fha,Teaney:2000cw,Teaney:2001av,Huovinen:2001cy,Hirano:2001eu,Hirano:2002ds},
which led to the development of later sophisticated dynamical models based on hydrodynamics
including viscosity%
~\cite{Song:2007ux, Dusling:2007gi, Luzem:2008cw, Schenke:2010nt, Bozek:2010wt, Song:2010mg, Song:2011qa, Schenke:2011tv}
and hydrodynamic fluctuations~%
\cite{Young:2014pka,muraseD,HF,Bluhm:2018plm,Singh:2018dpk,Sakai:2020pjw}.
The large elliptic flow was observed also at LHC%
~\cite{Aamodt:2010pa, ALICE:2011ab, Chatrchyan:2011pb,ATLAS:2011ah},
and later,
higher-order anisotropic flows have been systematically measured
at RHIC~\cite{Adare:2011tg,Adam:2019woz}
and LHC~\cite{ALICE:2011ab,ATLAS:2012at,Chatrchyan:2012wg}.

To understand collective flow phenomena more comprehensively, the correlation of the anisotropic flow
has been studied through,
e.g., the factorisation ratio
which was
initially proposed as a function of the transverse momentum~\cite{Gardim:2012im}
and extended in the longitudinal direction~\cite{Khachatryan:2015oea}.
The longitudinal factorisation ratios are widely measured in experiments%
~\cite{Khachatryan:2015oea, Aaboud:2017tql, Huo:2017hjv,
Nie:2019bgd, ATLAS:2020sgl},
where the rapidity decorrelation is observed as the factorisation breakdown.
The longitudinal dynamics
carries more information on the model-specific fluctuations
that are independent of the geometric origin of the initial nucleon distributions,
and thus the longitudinal factorisation ratio
is one of the good measures for the model discrimination.
These factorisation ratios are
studied with various initialisation models%
~\cite{Bozek:2010vz, Bozek:2017qir, Schenke:2016ksl, Shen:2017bsr, Jia:2017kdq,
Behera:2020mol}
and longitudinal-fluctuation mechanisms~\cite{Xiao:2012uw, Pang:2014pxa,
Bozek:2015bna, Pang:2015zrq, HF, Pang:2018zzo, Wu:2018cpc, Sakai:2020pjw}.
Although these models exhibit the factorisation breakdown,
none has quantitatively described all the measurements,
including the centrality dependence, different harmonic orders and the collision energy dependence,
in a single model setup.
In our previous study with the fluctuating hydrodynamic model~\cite{Sakai:2020pjw},
we have shown the significance of the hydrodynamic fluctuations
in understanding the longitudinal dynamics
while the initial fluctuations of longitudinal profiles were missing there.
Hydrodynamic fluctuations are thermal fluctuations of the hydrodynamic description
whose power is determined by the fluctuation--dissipation relation~%
\cite{Landau:1959,Lifshitz:1980,Calzetta:1997aj,Kapusta:2011gt,Murase:2013tma,An:2019osr,Murase:2019cwc}.
Besides, the initial-state longitudinal fluctuations
also play an important role in the final-state decorrelations
\cite{Xiao:2012uw, Pang:2014pxa, Bozek:2015bna, Pang:2015zrq,
Pang:2018zzo, Wu:2018cpc}.
Therefore, in this Letter, we investigate both effects of
the longitudinal fluctuations in the initial stage and
the hydrodynamic fluctuations in the hydrodynamic stage
on the rapidity decorrelation.

In this Letter,
we employ the integrated dynamical model
with hydrodynamic fluctuations~\cite{Murase:2016rhl,Sakai:2020pjw},
where the causal fluctuating hydrodynamic code {\tt rfh}~\cite{muraseD}
is combined with the initialisation model and the cascade model {\tt JAM}~\cite{Nara:1999dz}
with the prescription described in Ref.~\cite{ppnp}.
For the initialisation model, we newly implement the initial longitudinal fluctuations
in the Monte-Carlo version of the Glauber model.
The constitutive equations for the shear-stress tensor, $\pi^{\mu\nu}$,
in the causal fluctuating hydrodynamics are chosen as
\cite{muraseD, Murase:2019cwc, Baier:2007ix}
\begin{align}
  \label{eq:shear_stress_tensor}
 \tau_\pi{\Delta^{\mu\nu}}_{\alpha\beta}u^{\lambda}\covDeriv_{;\lambda}\pi^{\alpha\beta}
  + \pi^{\mu\nu}\left(1+\frac{4}{3}\tau_\pi\covDeriv_{;\lambda}u^\lambda \right)
  = 2\eta{\Delta^{\mu\nu}}_{\alpha\beta}\covDeriv^{;\alpha} u^\beta+\xi^{\mu\nu},
\end{align}
where $\eta$ and $\tau_{\pi}$ are the shear viscosity and the relaxation time, respectively.
The tensor $\Delta^{\mu\nu} = g^{\mu\nu} - u^\mu u^\nu$
is a projector for four-vectors onto the components transverse to the flow velocity $u^\mu$
where the sign convention for the metric is $g_{\mu\nu} = \diag(+,-,-,-)$.
The tensor ${\Delta^{\mu\nu}}_{\alpha\beta}=\frac{1}{2}\left({\Delta^\mu}_\alpha{\Delta^\nu}_\beta+{\Delta^\mu}_\beta{\Delta^\nu}_\alpha \right)-\frac{1}{3}{\Delta^{\mu\nu}}\Delta_{\alpha\beta}$
is a projector for second-rank tensors onto the symmetric and traceless components transverse to the flow velocity.
The symbol $\covDeriv_{;\mu}$ denotes the covariant derivative.
The noise term $\xi^{\mu \nu}$ represents the hydrodynamic fluctuations
and obeys the fluctuation--dissipation relation.
In the Milne coordinates $(\tau,\eta_{\mathrm{s}},\bm{x}_\perp) := \bigl(\sqrt{t^2 - z^2},\tanh^{-1}(z/t),x,y\bigr)$,
the fluctuation--dissipation relation is written as
\begin{align}
  \label{eq:FD_Milne}
  \langle\xi^{\mu\nu}(\tau, \eta_{\mathrm{s}}, \bm{x}_\perp)\xi^{\alpha\beta}(\tau', \eta_{\mathrm{s}}', \bm{x}'_\perp)\rangle
  = 4\eta T\Delta^{\mu\nu\alpha\beta}
  \cdot \frac1{\tau}\delta(\tau-\tau')
  \delta(\eta_{\mathrm{s}}-\eta_{\mathrm{s}}')\delta^{(2)}(\bm{x}_\perp - \bm{x}_\perp'),
\end{align}
where $T$ is the temperature, and
the angle brackets mean ensemble average,
which is associated with the event average
in the context of the high-energy nuclear collisions.
Here, the Lorentz indices in Eq.~\eqref{eq:FD_Milne}
represent $\tau$, $\eta_{\mathrm{s}}$, $x$ or $y$.
In the actual calculations,
we introduce the spatial cutoff to the hydrodynamic fluctuations
by convoluting the Gaussian profile of widths $\lambda_\eta$ and $\lambda_\perp$
in the longitudinal and transverse directions, respectively.
This effectively replaces the spatial delta function
in Eq.~\eqref{eq:FD_Milne}
by the Gaussian of width 2$\lambda_\eta$ and 2$\lambda_\perp$.
Smaller cutoff parameters result in larger effects of the hydrodynamic fluctuations.
It is noted here that these cutoff parameters are specified
in the coordinate space rather than in the momentum space in this Letter.

In the hydrodynamic simulations,
the initial entropy density distributions, $s(\tau_0, \eta_\mathrm{s}, \bm{x}_\perp)$,
are needed at a fixed initial time $\tau_0$.
For the event-by-event initial profiles, we utilise
the Monte-Carlo version of the Glauber (MC-Glauber) model~\cite{GLAUBER20063, ppnp}
combined with a general-purpose event generator, \pythia~\cite{Sjostrand:2014zea}.
For each binary collision,
we generate hadrons in a p+p collision using \pythia{}.
Here, we neglect the difference between p+p and p+n / n+n binary collisions
that happen in real nuclear collisions for simplicity.
If we would simply sum up all the hadrons from the binary collisions,
the multiplicity would scale with the number of the binary collisions, $N_{\mathrm{coll}}$,
which would be plausible only in high-transverse-momentum (high-$\pT$) regions.
On the other hand, yields of low-transverse-momentum (low-$\pT$) hadrons
are expected to scale with the number of participants, $N_{\mathrm{part}}$.
To embody this scaling behaviour throughout the whole transverse momentum regions,
we perform a rejection sampling for these generated hadrons
with the momentum-dependent acceptance
probability $w(Y, \pT)$~\cite{Okai:2017ofp, Kawaguchi:2017tuj}:
\begin{align}
\label{eq:acceptance_prob_YpT}
w(\pT, Y)
&= w(Y) \times \frac{1}{2} \left[1-\tanh \left(\frac{\pT-p_{\mathrm{T0}}}{\Delta \pT} \right) \right] 
+ \frac{1}{2} \left[1+\tanh \left(\frac{\pT-p_{\mathrm{T0}}}{\Delta \pT}\right)\right],\\
\label{eq:acceptance_prob_Y}
w(Y) & = \frac{Y_{\mathrm{b}}+Y}{2Y_\mathrm{b}}\frac{1}{n_\mathrm{A}} + \frac{Y_\mathrm{b}-Y}{2Y_\mathrm{b}}\frac{1}{n_\mathrm{B}}.
\end{align}
Here, $\pT$ and $Y$ are the transverse momentum and the rapidity of the hadron, respectively.
We introduce the parameters $p_{\mathrm{T0}}$ and $\Delta \pT$ to
smoothly separate low- and high-$\pT$ regions
into the first and second terms, respectively,
so that the total number of accepted hadrons scales
with $N_\mathrm{part}$ ($N_\mathrm{coll}$) in the low- (high-) $\pT$ region.
The symbol $Y_\mathrm{b}$ denotes the beam rapidity,
and $n_\mathrm{A}$ ($n_\mathrm{B}$) is the number of binary collisions
that the nucleon of the current binary collision at positive (negative)
beam rapidity experiences.
The scaling with $N_{\mathrm{part}}$ is implemented by the function $w(Y)$.
This function $w(Y)$ also brings rapidity-dependent yields of hadrons%
~\cite{Brodsky:1977de, Adil:2005qn, Hirano:2005xf},
which shares the idea with that of the wounded nucleon model \cite{Bialas:1976ed}.
The free parameters $p_{\mathrm{T0}}$ and $\Delta \pT$ are later
tuned to reproduce the centrality dependence of the charged-particle multiplicity.

We assume the initial entropy-density distribution is proportional
to the number distribution of the accepted hadrons:
\begin{align}
\label{eq:initial}
s(\tau_0, \eta_\mathrm{s}, \bm{x}_\perp) = \frac{K}{\tau_0} \sum_{i} \frac{1}{\sqrt{2\pi\sigma^2_\eta}}\frac{1}{2\pi\sigma^2_\perp}
 \exp\left[-\frac{\left(\bm{x}_\perp-\bm{x}_\perp^i \right)^2}{2\sigma^2_\perp}-\frac{\left( \eta_{\mathrm{s}}-\eta^i_{\mathrm{s}}\right)^2}{2\sigma^2_\eta}\right],
\end{align}
where the model parameter $K$ controls the overall normalisation.
The longitudinal position of the $i$-th hadron at the initial time $\tau_{0}$
is determined as $\eta_{\mathrm{s}}^i =Y^i$.
The initial transverse position of the $i$-th hadron $\bm{x}_\perp^i$ is
randomly sampled in the uniform disk of the radius $\sqrt{\sigma^\mathrm{NN}_\mathrm{in}/\pi}$
and the centre $\bm{x}_\mathrm{centre}^i$.
Here, $\sigma^\mathrm{NN}_\mathrm{in}(\sNN)$
is the inelastic-scattering cross section of nucleons,
and we place the centre at
\begin{align}
\bm{x}_\mathrm{centre}^i = \frac{\bm{x}_\perp^{i,\mathrm{A}} + \bm{x}_\perp^{i,\mathrm{B}}}{2}
+\frac{\bm{x}_\perp^{i,\mathrm{A}} - \bm{x}_\perp^{i,\mathrm{B}}}{2Y_\mathrm{b}}\eta_{\mathrm{s}}^i,
\end{align}
where $\bm{x}_\perp^{i,\mathrm{A}}$ ($\bm{x}_\perp^{i,\mathrm{B}}$) is the position of the associated nucleon
at the positive (negative) beam rapidity.
Here we assumed that a hadron is produced around a line in the Milne coordinates
connecting the two nucleons of the associated binary collision.

We put the Bjorken scaling solution~\cite{Bjorken:1982qr}
for the initial flow velocity $u^\mu(\tau_0,\eta_\mathrm{s},\bm{x}_\perp)
=(u^{\tau}, u^{\eta_\mathrm{s}}, u^x, u^y) = (1, 0, 0, 0)$,
which means that we ignore the fluctuations of initial flows in this study.
For the equation of state, we employ a lattice-based model, $s95p$-v1.1~\cite{Huovinen:2009yb},
in which the list of hadrons in the hadron resonance gas model is taken from the cascade model {\tt JAM}~\cite{Nara:1999dz}.
At the switching temperature $T_{\mathrm{sw}}$, we change the description from the macroscopic hydrodynamics to the microscopic kinetic theory using the Cooper--Frye formula~\cite{Cooper:1974mv}.
The subsequent space-time evolution of hadrons is described by using the cascade model {\tt JAM}~\cite{Nara:1999dz}.
Further details of the integrated dynamical model with fluctuating hydrodynamics
can be found in Refs.~\cite{muraseD,Murase:2019cwc,Sakai:2020pjw}.


Using this model,
we perform simulations for Pb+Pb collisions at $\sNN=2.76\ \text{TeV}$.
For the reference setup to see the effects of hydrodynamic fluctuations,
we also perform the simulations of viscous hydrodynamics
by turning off the hydrodynamic fluctuations.
For each hydrodynamic model (fluctuating hydrodynamics and viscous hydrodynamics),
we gain 4\;000 minimum-bias hydrodynamic events.
For each hydrodynamic event, we perform 100 independent particlisation and
hadronic cascade simulations to reduce the computational cost.
In total we obtain the 400\;000 $(=4\;000 \times 100)$ events for the subsequent analyses.

Let us here summarise the parameters of the present study.
Following the previous calculations~\cite{Murase:2016rhl,ppnp},
we set the specific shear viscosity $\eta/s = 1/4\pi$~\cite{Kovtun:2004de},
the relaxation time $\tau_{\pi} = 3/4\pi T$~\cite{Song:2009gc,Baier:2007ix},
the initial proper time $\tau_0 = 0.6\ \text{fm}$
and the switching temperature $T_{\mathrm{sw}}=155\ \text{MeV}$.
We fix the widths of the hadron profile as $\sigma_\perp = 0.3\ \text{fm}$ and
$\sigma_\eta = 0.3$ at present.
We tune initial parameters $K$, $p_{\mathrm{T0}}$ and $\Delta \pT$
for each hydrodynamic model
to reproduce centrality dependence of charged-particle multiplicity
measured by the ALICE Collaboration~\cite{Aamodt:2010cz}.
For the fluctuating hydrodynamics,
we tune the cutoff parameters $\lambda_\perp$ and $\lambda_\eta$
to roughly reproduce the factorisation ratio $r_2(\etapa, \etapb)$
measured by the CMS Collaboration~\cite{Khachatryan:2015oea}.
In the present study we assume $\lambda_\perp/\text{fm} = \lambda_\eta$ for simplicity.
These parameters for each hydrodynamic model
are summarised in Table~\ref{table:parameter}.
\begin{table}[htbp]
\caption{Parameters in hydrodynamic models}
\begin{tabular}{lccccc} \hline \hline
Model & $\lambda_\perp$ (fm) & $\lambda_\eta$ & $K$ & $p_{\mathrm{T0}}$ (GeV) & $\Delta \pT$ (GeV)  \\ \hline
Viscous hydro            & N/A & N/A & 4.8 & 1.75 & 1.0 \\
Fluc. hydro-$\lambda$2.0 & 2.0 & 2.0 & 4.8 & 1.80 & 1.0 \\ \hline \hline
\end{tabular}
\label{table:parameter}
\end{table}
The parameter $K$ controls the overall magnitude
of multiplicity per participant pair.
On the other hand, the parameter $p_{\mathrm{T0}}$ controls not only the overall magnitude
but also the slope of multiplicity per participant pair.
The parameter $p_{\mathrm{T0}}$ is slightly larger in fluctuating hydrodynamics
because the entropy production in the hydrodynamic stage in the peripheral collisions
is larger with fluctuating hydrodynamics than with viscous hydrodynamics~\cite{Sakai:2020pjw}.

\begin{figure}[htbp]
\begin{center}
\centering
\includegraphics[width=0.5\textwidth, bb=60 0 320 220]{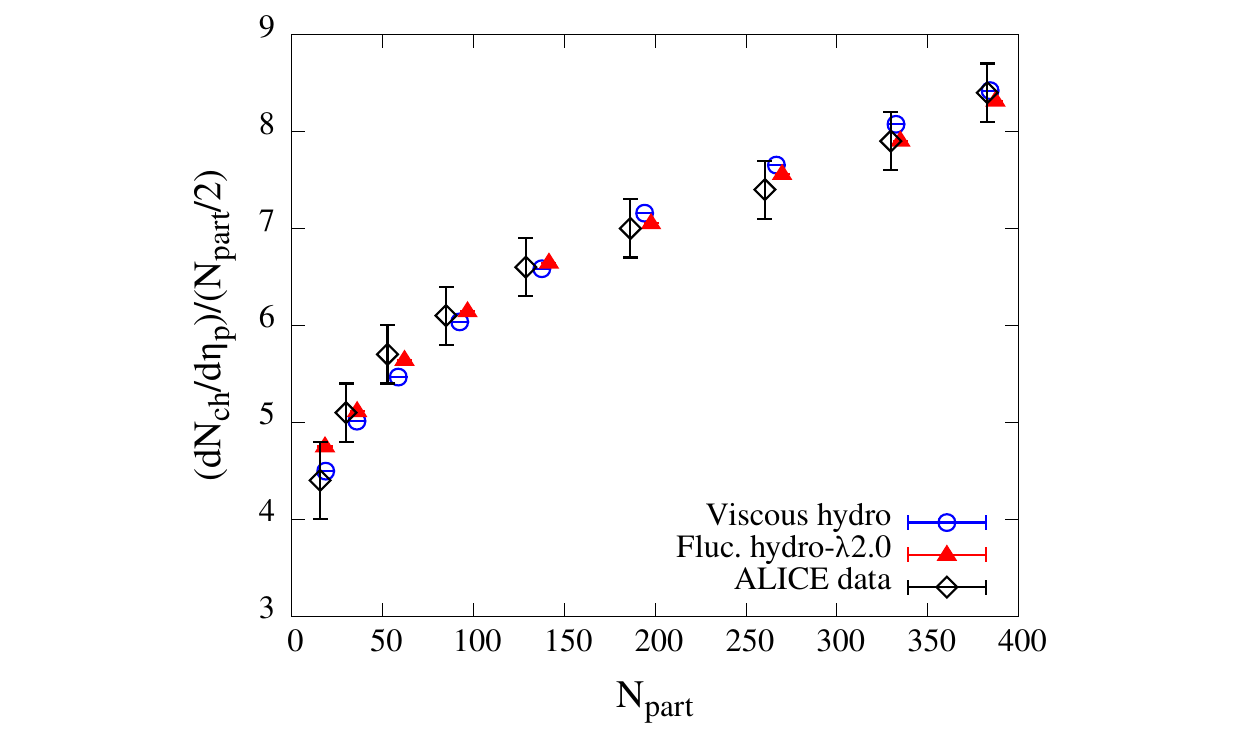}
\caption{(Colour Online) Charged-hadron multiplicity
normalised by the number of the participant pair
$(dN_\mathrm{ch}/d\etap)/(N_\mathrm{part}/2)$,
as a function of the number of participants.
Results from viscous hydrodynamics with initial longitudinal fluctuations (open circle)
and fluctuating hydrodynamics with initial longitudinal fluctuations (filled triangle)
are compared with experimental data (open diamond)
obtained by the ALICE Collaboration~\cite{Aamodt:2010cz}.
}
\label{fig:multiplicity}
\end{center}
\end{figure}

Figure~\ref{fig:multiplicity} shows $N_\mathrm{part}$ dependence of charged-hadron multiplicity
$dN_\mathrm{ch}/d\etap$ per participant pair $N_\mathrm{part}/2$ at midrapidity $|\etap|<0.5$
in Pb+Pb collisions at $\sNN=2.76\ \text{TeV}$.
Within our initialisation model,
the violation of $N_\mathrm{part}$ scaling in multiplicity in Fig.~\ref{fig:multiplicity}
is described by the initial (semi-)hard components introduced through
the acceptance probability Eq.~\eqref{eq:acceptance_prob_YpT}.


The initial longitudinal fluctuations and the hydrodynamic fluctuations,
which follow Eqs.~\eqref{eq:shear_stress_tensor} and \eqref{eq:initial}, respectively,
randomly disturb the correlations such as
alignment of the event planes along rapidity.
We analyse factorisation ratios~\cite{Gardim:2012im}
to see the effects of these fluctuations on
(de-)correlation of event-plane angles along rapidity.
The factorisation ratio in the longitudinal direction is defined as
\begin{align}
\label{eq:factorisation_ratio}
r_n(\etapa, \etapb) = \frac{V_{n\Delta}(-\etapa, \etapb)}{V_{n\Delta}(\etapa, \etapb)}, \quad
V_{n\Delta} = \langle{\cos (n\Delta\phi)} \rangle.
\end{align}
Here $V_{n\Delta}$ is the Fourier coefficients of two-particle correlation functions
at the $n$-th order and
 $\Delta\phi$ is a difference of the azimuthal angles between two charged hadrons in separated pseudorapidity regions,
$\etapa$ and $\etapb$.
If one could factorise the two-particle correlation functions
in the denominator and numerator in Eq.~\eqref{eq:factorisation_ratio}
into two anisotropic flows with corresponding momentum regions,
e.g., $V_{n\Delta} (\etapa, \etapb) = v_2(\etapa)v_2(\etapb)$,
the resultant factorisation ratio would be equal to unity in symmetric collisions.
Contrarily, when the event-plane angle and the flow magnitude fluctuate as functions of pseudorapidity,
the ratio becomes smaller than unity in general
because one cannot factorise the two-particle correlation functions.
This decrease of the factorisation ratio from the unity
is called the flow decorrelation in the longitudinal direction.
In the following analyses,
the two-particle correlation function
is calculated by changing the rapidity region of the hadrons
within $0<\etapa<2.5$ while fixing the region of the reference hadrons
to be $3.0<\etapb<4.0$
following the experimental setup by the CMS Collaboration~\cite{Khachatryan:2015oea}.

\begin{figure}[htbp]
\centering
\includegraphics[width=0.48\textwidth, bb=60 0 320 220]{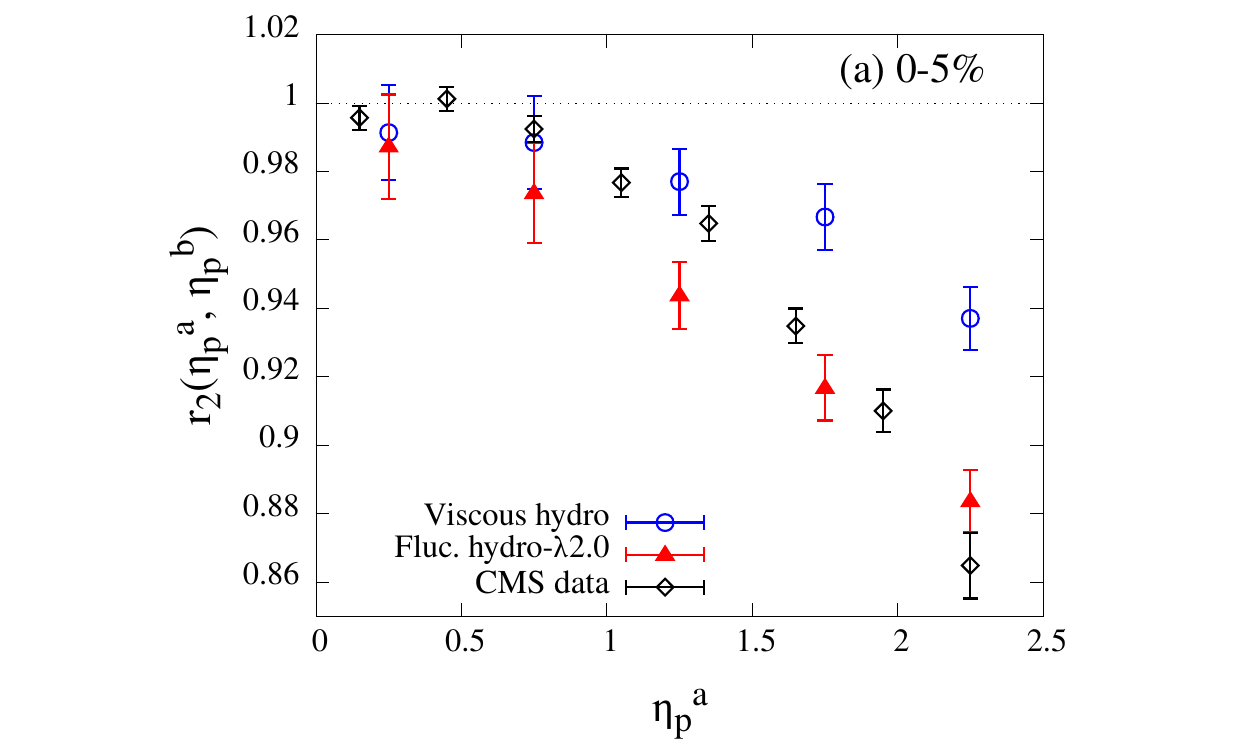}
\includegraphics[width=0.48\textwidth, bb=60 0 320 220]{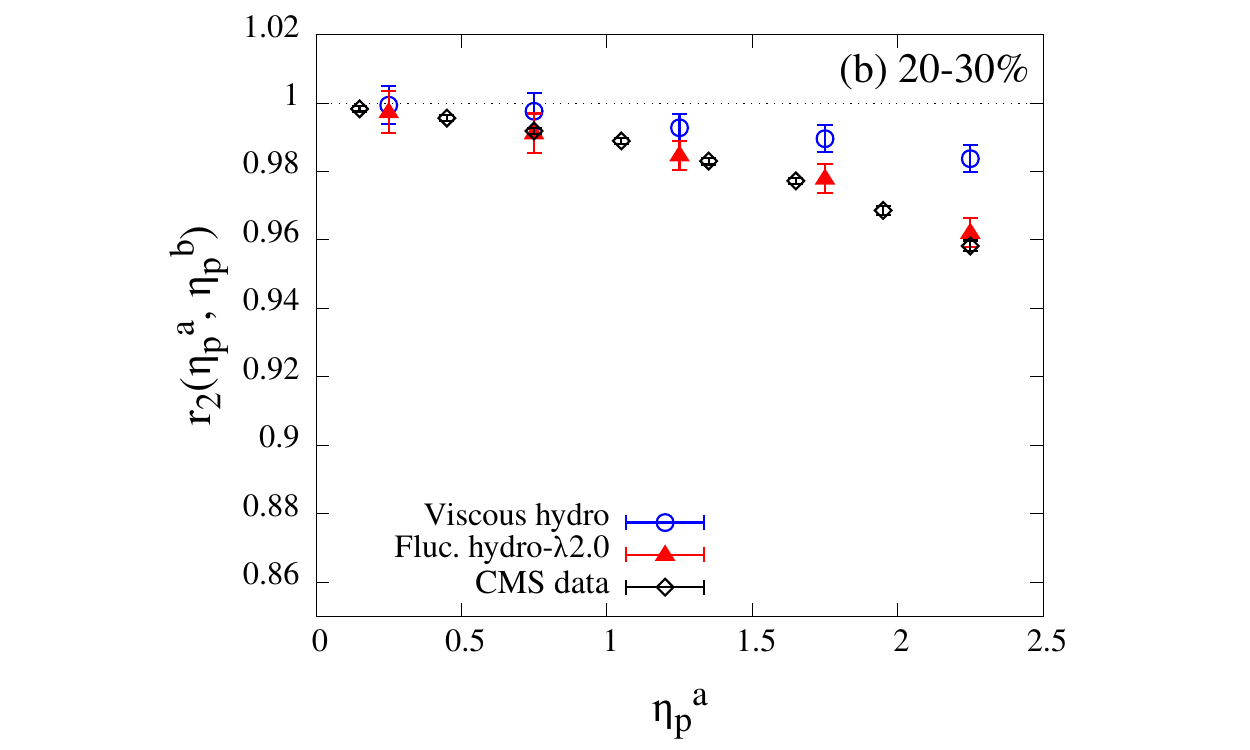}
\caption{Factorisation ratio $r_2(\etapa,\etapb)$ in Pb+Pb collisions
at $\sNN = 2.76$ TeV for (a) 0--5\% and (b) 20--30\%
centralities.
The rapidity region for reference is $3.0<\etapb<4.0$.
Open circles and filled triangles
are results from the viscous and fluctuating hydrodynamic models, respectively.
The experimental data of $r_2(\etapa, \etapb)$
from the CMS Collaboration~\cite{Khachatryan:2015oea} are shown by open diamonds.
}
\label{fig:r2}
\end{figure}

Figure~\ref{fig:r2} shows the factorisation ratio in the longitudinal direction, $r_2(\etapa, \etapb)$,
in Pb+Pb collisions at $\sNN = 2.76$ TeV for 0--5\% and 20--30\% centralities
compared with the experimental data
by the CMS Collaboration~\cite{Khachatryan:2015oea}.
The cutoff parameters $\lambda_\perp/\text{fm} = \lambda_\eta = 2.0$ in Table~\ref{table:parameter}
have been here determined to reproduce the experimental results of $r_2(\etapa,\etapb)$ of the 20--30\% centrality.
These values are larger than $\lambda_\perp/\text{fm} = \lambda_\eta = 1.0$ and $1.5$
(parameters $\lambda1.0$ and $\lambda1.5$, respectively)
of the previous work~\cite{Sakai:2020pjw}.
This is because
the effect of newly introduced initial longitudinal fluctuations
needs to be compensated by reducing the effect of the hydrodynamic fluctuations
to keep the factorisation ratios the same.

The experimental data
is close to unity at small $\etapa$ and decreases with increasing $\etapa$.
The factorisation ratio $r_2(\etapa, \etapb)$ from the viscous hydrodynamic model
with initial longitudinal fluctuations decreases with increasing $\etapa$.
However, the decorrelation is weaker than experimental data in particular at
the large rapidity gap $\etapa\sim 1.5$--$2.5$.
Since viscous hydrodynamics tends to keep the long-range correlation
in the rapidity direction~\cite{Sakai:2020pjw},
the decreasing behaviour of $r_2(\etapa, \etapb)$
can be attributed to the initial longitudinal fluctuations~%
\cite{Wu:2018cpc, Bozek:2015bna,
Pang:2014pxa, Pang:2015zrq, Pang:2018zzo}.
In the fluctuating hydrodynamic model,
the decorrelation is stronger than in the viscous hydrodynamic model,
which is the same trend as in the previous analysis
without the initial longitudinal fluctuations in Ref.~\cite{Sakai:2020pjw}.

\begin{figure*}[htbp]
\begin{center}
\centering
\includegraphics[width=0.49\textwidth, bb=55 0 325 220]{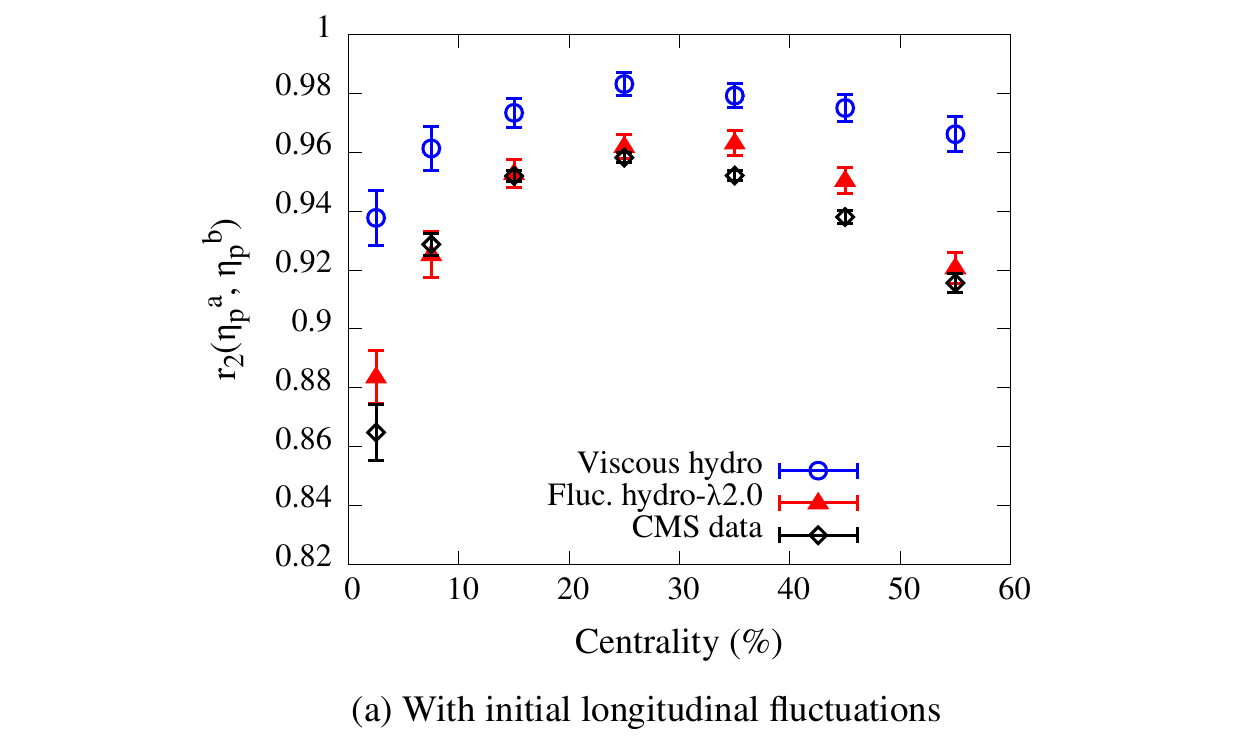}
\includegraphics[width=0.49\textwidth, bb=55 0 325 220]{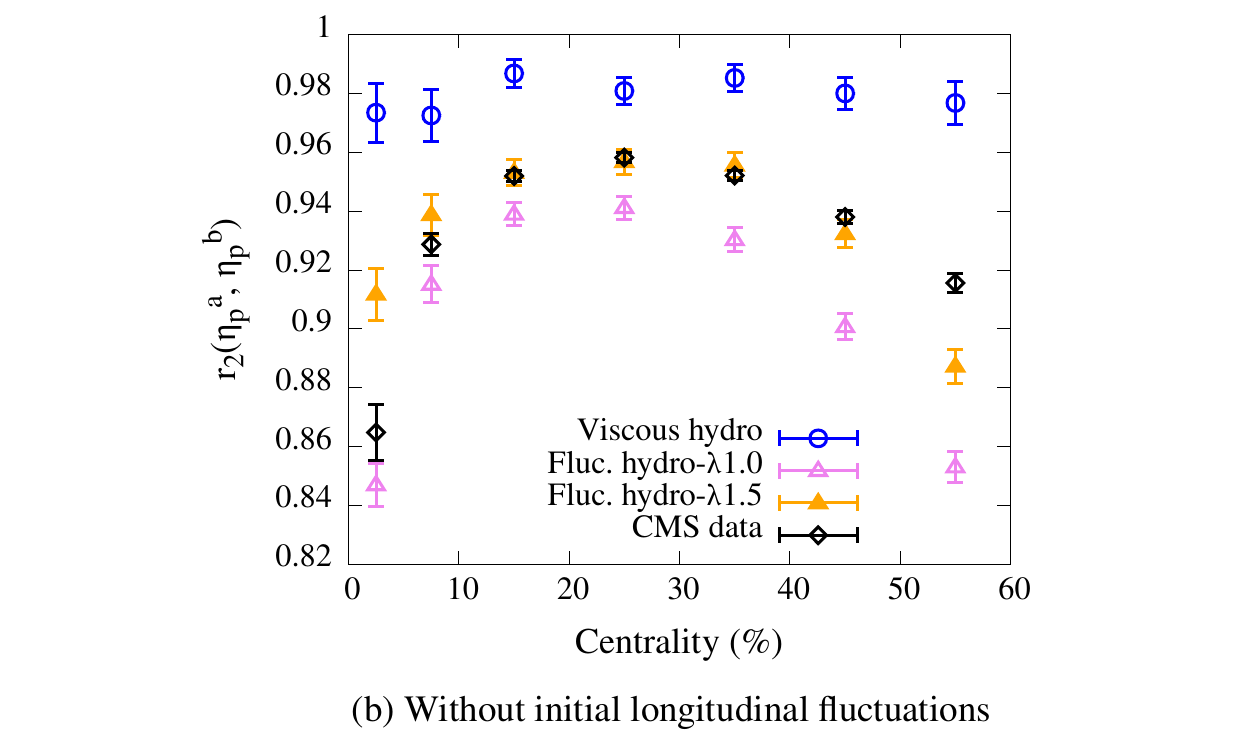}
\caption{(Colour Online)
Centrality dependence of factorisation ratio $r_2(\etapa, \etapb)$
in Pb+Pb collisions at $\sNN = 2.76$ TeV.
The rapidity regions of two-particle correlation functions are
taken as $2.0<\etapa<2.5$ and $3.0<\etapb<4.0$.
(a) The results from hydrodynamic models with initial longitudinal fluctuations
are compared with the experimental data~\cite{Khachatryan:2015oea}.
(b) The results from hydrodynamic model
without initial longitudinal fluctuations (taken from Fig.~6 in Ref.~\cite{Sakai:2020pjw})
are also shown for comparison.
In the right panel (b), we show the result from viscous hydrodynamics
and two results from the fluctuating hydrodynamics
with different cutoff parameters $\lambda1.5$ (filled triangle) and
$\lambda1.0$ (open triangle),
which correspond to the cutoff $\lambda_\perp/\text{fm} = \lambda_\eta = 1.5$ and $1.0$, respectively
\relax[see Ref.~\cite{Sakai:2020pjw} for the other parameters used in panel (b)].
The symbols are the same as in Fig.~\ref{fig:r2} for other plots.
}
\label{fig:r2comp}
\end{center}
\end{figure*}

Figures~\ref{fig:r2comp} (a) and (b) compare
the centrality dependence of $r_2(\etapa, \etapb)$
from hydrodynamic models with
and
without initial longitudinal fluctuations, 
respectively.
We find that the experimental data of the centrality dependence of $r_2(\etapa, \etapb)$
can only be reproduced by the model with
both the hydrodynamic fluctuations and the initial longitudinal fluctuations.
This can be understood from the different centrality dependence between the effects
of the initial longitudinal fluctuations and the hydrodynamic fluctuations.

Within our model, the genuine initial longitudinal fluctuations tend
to decrease $r_2(\etapa, \etapb)$ in central collisions (particularly in 0--10\% centrality),
as seen
in the open circles
in Fig.~\ref{fig:r2comp} (a).
This is because the flow correlation along the rapidity direction
can be easily broken in the central collisions
as the magnitude of the elliptic flow driven by the collision geometry
is small there.
On the other hand,
the genuine hydrodynamic fluctuations tend to decrease $r_2(\etapa, \etapb)$
in both central collisions (0--10\% centrality) and peripheral collisions (40--60\% centrality)
as seen 
in Fig.~\ref{fig:r2comp} (b).
The mechanism of the decorrelation in central collisions is the same as with the initial longitudinal fluctuations.
The decorrelation in the peripheral collisions can be attributed to the nature of
the hydrodynamic fluctuations being significant in small and short-lived systems,
i.e., the magnitude of the averaged thermal fluctuations at the linear order scales as $(V t)^{-1/2}$
where $V$ and $t$ are the typical volume and lifetime of the system.
If we do not take account of initial longitudinal fluctuations,
the factorisation ratio $r_2(\etapa, \etapb)$ of the experimental data
can be well fitted by parameter $\lambda1.0$ (open triangles) and $\lambda1.5$ (filled triangles)
in central collisions and peripheral collisions, respectively,
but the overall centrality dependence
cannot be reproduced by a single cutoff parameter.

Neither the fluctuating hydrodynamics without the initial longitudinal fluctuations with fixed $\lambda$ parameters
nor the viscous hydrodynamics with the initial longitudinal fluctuations  can
reproduce the centrality dependence of the experimental data.
Thus we conclude that \textit{both the initial longitudinal fluctuations
and the hydrodynamic fluctuations} must be taken into consideration
in understanding the centrality dependence of the factorisation ratios.

\begin{figure}[htbp]
\centering
\includegraphics[width=0.48\textwidth, bb=55 0 325 220]{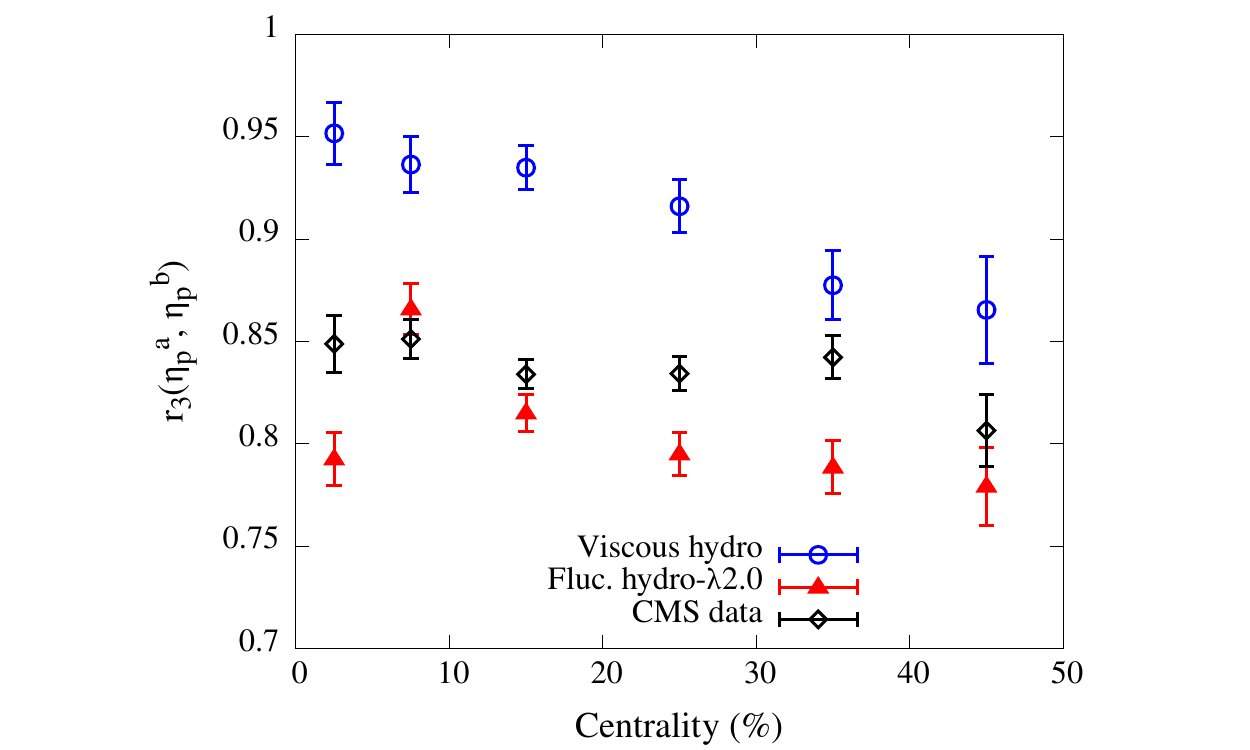}
\caption{(Colour Online)
Centrality dependence of factorisation ratio $r_3(\etapa, \etapb)$
in Pb+Pb collisions at $\sNN = 2.76$ TeV.
The rapidity regions of two-particle correlation functions are
taken as $2.0<\etapa<2.5$ and $3.0<\etapb<4.0$.
The symbols are the same as in Fig.~\ref{fig:multiplicity}.
The experimental data are obtained by the CMS Collaboration~\cite{Khachatryan:2015oea}.
}
\label{fig:r3comp}
\end{figure}

Figure~\ref{fig:r3comp} shows the centrality dependence of $r_3(\etapa, \etapb)$
in Pb+Pb collisions at $\sNN = 2.76$ TeV.
The factorisation ratio $r_3(\etapa, \etapb)$
in experimental data depends mildly on centrality.
This is because
the triangular flow is dominated by fluctuations
in symmetric collision systems.
The factorisation ratio $r_3(\etapa, \etapb)$ from the viscous hydrodynamic model
with initial longitudinal fluctuations
decreases with increasing centrality percentage
and is larger than the experimental data in all centralities.
In contrast,
$r_3(\etapa, \etapb)$ obtained from the fluctuating hydrodynamic model
has less centrality dependence
and is also close to experimental data.
Although more sophisticated modelling would be required to
perfectly reproduce the experimental data
of both $r_2(\etapa, \etapb)$ and $r_3(\etapa, \etapb)$ simultaneously,
the centrality dependences
of $r_2(\etapa, \etapb)$ and $r_3(\etapa, \etapb)$
are better reproduced by including both the hydrodynamic fluctuations and
the initial longitudinal fluctuations.
These results suggest that
considering either the initial longitudinal fluctuations
or the hydrodynamic fluctuations is insufficient
in understanding the decorrelation of anisotropic flows in the longitudinal directions
and thus that considering both simultaneously in a model is important.

In this Letter, we investigated the effects
of the longitudinal fluctuations in the initial stage
and the hydrodynamic fluctuations in the expansion stage
on the longitudinal factorisation ratios
to understand the rapidity decorrelation of anisotropic flows.
We employed an integrated dynamical model
which consists of
the initialisation model for fluctuations in both longitudinal
and transverse profiles,
the hydrodynamic model, {\tt rfh},
for causal hydrodynamic fluctuations and dissipations,
and the hadronic cascade model, {\tt JAM},
for the final-state interactions.
We performed simulations of Pb+Pb collisions at $\sNN=2.76\ \text{TeV}$
using the integrated dynamical model with or without
the hydrodynamic fluctuations for comparison.
We fixed the model parameters so that our model fairly
reproduces the experimental data of the centrality dependence of 
charged-hadron multiplicity normalised by the number of participants.
With these model settings,
we calculated the factorisation ratio $r_n(\etapa, \etapb)$ $(n=2, 3)$
in the longitudinal direction
and its centrality dependence.
We confirmed that both the initial longitudinal fluctuations
and the hydrodynamic fluctuations decrease the factorisation ratio, which is
consistent with the previous studies.
However, we found that the centrality dependence of the effects of hydrodynamic fluctuations is different from those of initial longitudinal fluctuations.
We reproduce the centrality dependence of the factorisation ratio
at the second order, $r_2(\etapa, \etapb)$, from the CMS collaboration
only after including both
the initial longitudinal and the hydrodynamic fluctuations.
We also qualitatively described
the centrality dependence of the factorisation ratio
at the third order,
$r_3(\etapa, \etapb)$, with these two fluctuations.
These analyses show that
incorporating both the initial longitudinal and the hydrodynamic fluctuations
simultaneously in a dynamical model
is the key to quantitatively understanding
the decorrelation dynamics in the longitudinal direction.

\section*{Acknowledgement}
This work was supported by JSPS KAKENHI Grant Numbers JP18J22227 (A.S.) and JP19K21881 (T.H.).
K.M. was supported by the NSFC under Grant No. 11947236.



\bibliographystyle{elsarticle-num} 
\bibliography{References}





\end{document}